\documentclass[twocolumn]{book}
\usepackage{graphicx,color}
\usepackage{makeidx,universe}

\makeauthorindex

\BookTitle{Proceedings of the XXIX PHYSICS IN COLLISION, Fermilab Preprint\# FERMILAB-CONF-09-498-E}

\CopyRight{\copyright 2009 by Universal Academy Press, Inc.}

\begin{document} 

\pagenumbering{arabic}

\chapter{%
{\LARGE \sf
Experimental probes of axions } \\
{\normalsize \bf 
Aaron S. Chou  } \\
{\small \it \vspace{-.5\baselineskip}
 Fermi National Accelerator Laboratory, 
      Wilson and Kirk Roads, Batavia, IL 60510-5011, U.S.A 
}
}


\AuthorContents{A.S.\ Chou}

\AuthorIndex{Chou}{A.S.}

  \baselineskip=10pt 
  \parindent=10pt    

\section*{Abstract} 

Experimental searches for axions or axion-like particles rely on semiclassical phenomena resulting from the postulated coupling of the axion to two photons.  Sensitive probes of the extremely small coupling constant can be made by exploiting familiar, coherent electromagnetic laboratory techniques, including resonant enhancement of transitions using microwave and optical cavities, Bragg scattering, and coherent photon-axion oscillations.  The axion beam may either be astrophysical in origin as in the case of dark matter axion searches and solar axion searches, or created in the laboratory from laser interactions with magnetic fields.  This note is meant to be a sampling of recent experimental results.

\section{Introduction} 
Axion models are motivated by the strong CP problem--the apparent vanishing of the CP- and T-violating electric dipole moment (EDM) of the neutron.  The total EDM is expected to receive contributions from both the TeV electroweak scale, via the quark spin, and from the GeV QCD scale, via the spatial distribution of the quark wavefunction within the neutron.  It is difficult to understand how these two contributions could cancel to such precision to produce a CP-conserving QCD ground state without fine-tuning of parameters.

The axion model of Peccei, Quinn, Weinberg, and Wilczek \cite{Peccei:1977ur, Peccei:1977hh, Wilczek:1977pj,Weinberg:1977ma} offers a dynamical solution to the strong CP problem by introducing a new scalar field which rolls within its potential into a state of minimum action, a CP-conserving QCD vacuum state.  Any imbalance between the contributions to the EDM from the TeV and GeV scales is absorbed into the scalar field value.  The quantized excitations of the scalar field about the potential minimum are called axions.  

The complex scalar potential starts as a Higgs-like Mexican hat potential, with symmetry-breaking scale $f$.  A massless Goldstone boson lives in the circular minimum of this potential at radius $f$ in field space.  During the QCD phase transition, instanton effects give a linear tilt to this potential, lifting it by an amount $\Lambda_\mathrm{QCD}^4$ on one side.  The degeneracy of the circular minimum is lifted, and the Goldstone boson starts rolling towards its minimum.  While rolling, a portion of the energy from the quark-gluon plasma is stored temporarily as potential energy.  The quantized excitations about the potential minimum are called axions, and have mass 
\begin{equation} 
\label{E:mass}
m_a \approx \Lambda_\mathrm{QCD}^2/f.
\end{equation}
When the minimum is reached, the original potential energy can be released as quanta of cold axions which could make up all or part of the inferred cold dark matter density of the universe.  

The usually quoted search window for axions is $10^{-6}\mbox{ eV} < m_a < 10^{-2}\mbox{ eV}$.  Axions with $m_a>10^{-2}\mbox{ eV}$ are disfavored by the supernova SN1987A cooling rate via the direct coupling of axions to fermions, suppressed by $1/f$, which is then linked to mass via Eq.~\ref{E:mass}.  Axions with $m_a<10^{-6}\mbox{ eV}$ may remain cosmologically frozen for too long before the potential energy density is converted into quanta, thus overproducing the cold dark matter density.  However, the total initial potential energy depends on the random QCD misalignment angle which describes where the original Goldstone boson was sitting in its circular potential before the potential was tilted.  If the field was originally already sitting close to the eventual minimum of the potential, then total amount of potential energy available for conversion into dark matter is restricted, and lower values of $m_a$ are allowed.

The natural coupling of axions $a$ to gluons or quarks is transferred to a coupling $g$ to the photon field strength $F$ via an anomaly diagram.
\begin{equation}
\label{E:interaction}
\mathcal{L} \approx \frac{\alpha}{8 \pi f} F\tilde{F}\equiv  -\frac{1}{4} g a F \tilde{F} = g a \vec{E}\cdot\vec{B}.
\end{equation}
This electromagnetic coupling represents the best experimental hope to discover axions.  For agnostic, model-independent searches, the experimental results are typically presented in the two-dimensional parameter space of photon coupling $g$ vs axion mass $m_a$.

\begin{figure}[t]
  \begin{center}
    \includegraphics[width=0.30\textwidth]{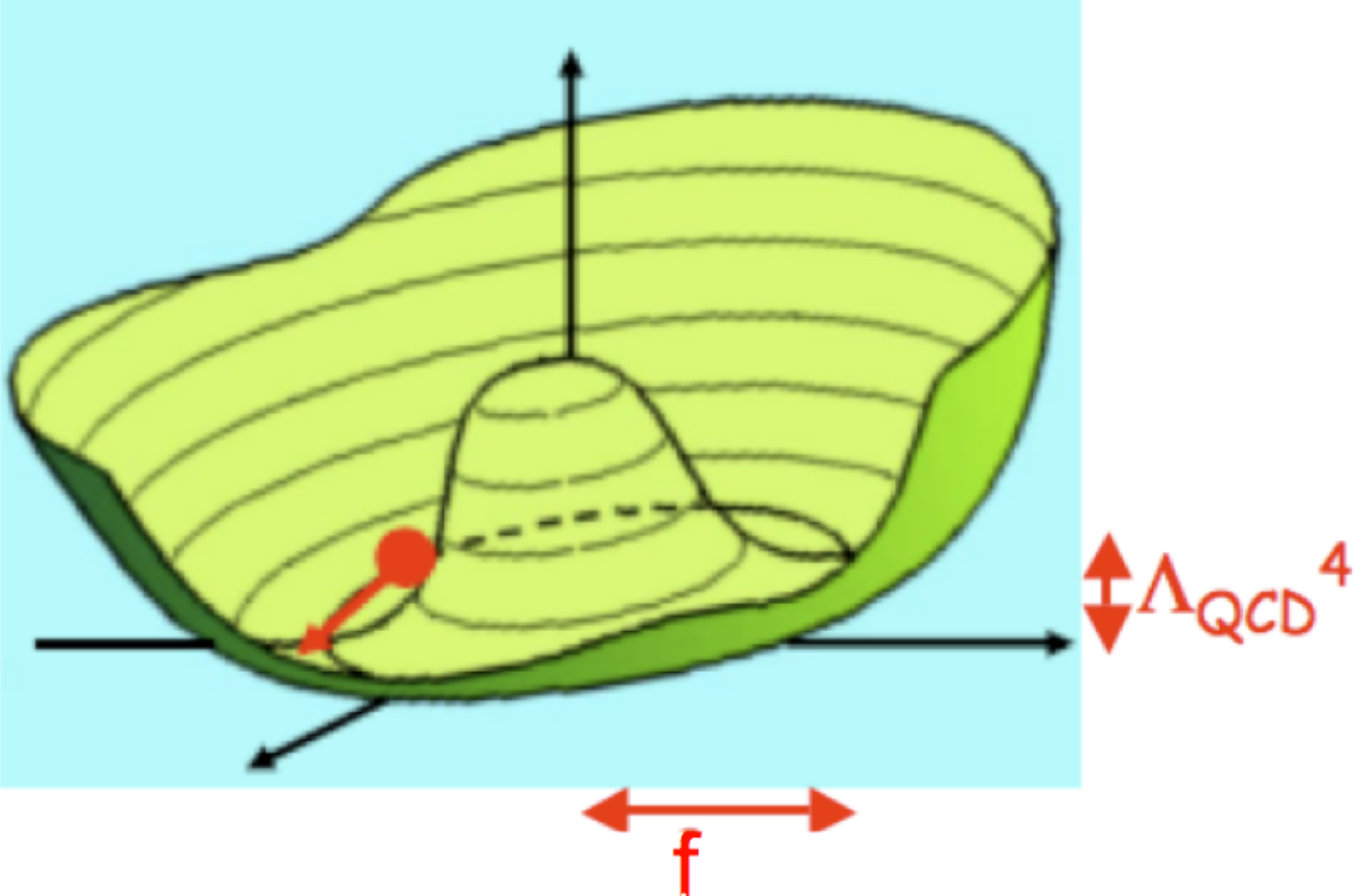}
  \end{center}
  \vspace{-1pc}
  \caption{The Peccei-Quinn symmetry is broken at some high mass scale $f$, resulting in a Higgs-like potential.  The potential is tilted by instanton effects during the QCD phase transition, causing the Goldstone boson to roll towards a unique minimum.  The quanta of small excitations about this minimum are called axions, and have mass $m_a=\Lambda_\mathrm{QCD}^2/f$.}
\end{figure}

\begin{figure}[t]
  \begin{center}
    \includegraphics[width=0.40\textwidth]{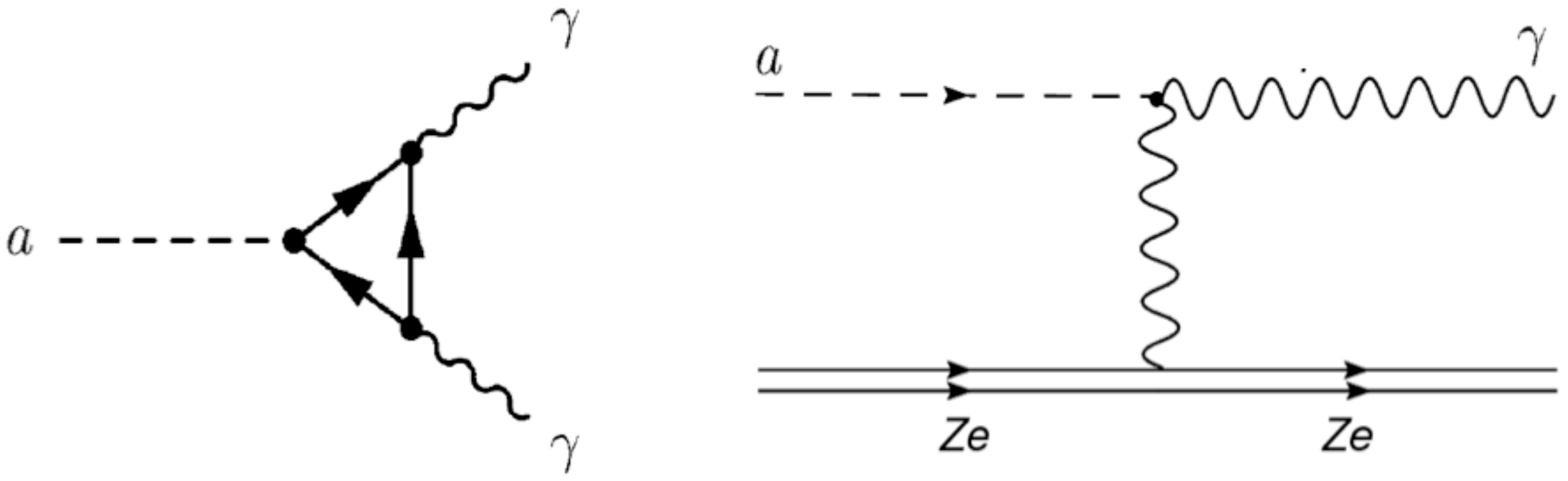}
  \end{center}
  \vspace{-1pc}
  \caption{(Left) While by design, axions are coupled to quarks and gluons, they also obtain couplings to two photons via triangle anomaly diagrams.  These effective couplings allow conversions between axions and photons, in the presence of background electromagnetic fields.  (Right) In Primakoff scattering, the axions convert into photons by interacting with the Coulomb field of a charged fermion.}
\end{figure}

\section{Axion Dark Matter Search (ADMX)} 
The ADMX experiment is based on the axion haloscope idea of Sikivie.  This idea is based on the modification to the Maxwell equations due to the photon-axion interaction.  In particular:
\begin{equation}
\vec{\nabla} \times \vec{B} - \partial_t\vec{E} = -g \vec B \partial_t a.
\end{equation}
In the presence of a constant background field $B_0$, the curl term vanishes, and the electric field follows the time evolution of the axion field, with amplitude suppressed by $g B_0$.  Due to the low dark matter halo velocity $v\approx 200$~km/s, the cold dark matter axions are expected to form a classical background field with coherence length greater than 100m, much larger than the size of a typical detector.  In the presence of a laboratory magnetic field, this spatially constant background axion field will drive an oscillating electric field with microwave frequency $E_a \approx m_a$.  Using a microwave cavity, this electric field can be built up coherently, with a quality factor limited by the intrinsic energy spread of the axions.  This kinetic energy spread is expected to be $\Delta E/E \approx 10^{-6}$ for halo axions which have thermalized into a Maxwellian distribution, or as small as $\Delta E/E\approx 10^{-11}$ for cold flows of dark matter which are entering the galaxy and have not yet thermalized.  

The ADMX experiment \cite{Duffy:2006aa,Duffy:2005ab,Bradley:2003kg} uses a meter-scale microwave cavity suffused with a 8 T field generated by a solenoidal magnet.  Cavity power at the level of $10^{-23}$~W can be detected using antennas amplified by HEMTs or, in a recent upgrade, by SQUID magnetometers.  The resonant frequency of the cavity can be tuned by inserting dielectric rods into the cavity.  The very small energy spread of the dark matter axions implies that millions of trials are necessary to scan the cavity frequency through a range of axion masses, thus limiting the integration time at each step in frequency to be on the scale of minutes.  So far, no excess power has been detected above thermal noise in any of the trials.  These results can be interpreted in two ways.  First, if axions are assumed to be the dominant component of cold dark matter, then this sets the normalization of the background axion field amplitude, and a non-observation of electric field gives an upper limit on the coupling $g$.  Alternatively, if the coupling $g$ is assumed to be that predicted by one of the popular models of axions such as the DFSZ or KSVZ model, then an upper bound can be set on the density of dark matter axions, which may contribute just a fraction of the total cold dark matter density.  The present results constrain the model in which KSVZ-type axions are the dominant form of dark matter.

Experimental scaling laws may be derived from the Dicke radiometer equation which specifies the signal to noise ratio
\begin{equation}
\frac{S}{N} = \frac{P_\mathrm{signal}}{kT_\mathrm{noise}} \sqrt{\frac{t}{\Delta \nu}}
\end{equation}
where $t$ is the integration time and $\Delta \nu$ is the frequency bandwidth, and the expected cavity signal power
\begin{equation}
P_\mathrm{signal} \propto \frac{g^2 B_0^2 Q  \rho_a  V }{m_a} 
\end{equation} 
where $Q$ is the limiting quality factor, $\rho_a$ is the axion dark matter energy density, and $V$ is the volume of the cavity.  For a fixed $S/N$ requirement, these two equations yield a scan rate
\begin{equation}
\frac{d\nu}{dt} \propto \frac{B_0^4 V^2}{T_\mathrm{noise}^2}
\end{equation}
for fixed sensitivity to coupling $g$, or a sensitivity to small couplings
\begin{equation}
g \propto \sqrt{\frac{T_\mathrm{noise}}{B_0^2 V}}
\end{equation}
for fixed scan rate.  The current activity in ADMX is to reduce the system noise temperature $T_\mathrm{noise}$.  The 2K HEMT-amplified sensors in the initial phase of the experiment have been replaced with SQUID magnetometers which currently have a noise temperature of 0.4K.  A further upgrade will add a dilution refrigerator to reduce the SQUID noise temperature to 0.1K.  This noise reduction will allow the scan rate to be increased by a factor of $\sim 250$, thus allowing more rapid coverage of the preferred range of masses for dark matter axions in the $10^{-6}\mbox{ eV}-10^{-5}\mbox{ eV}$ range.  Extending this technique towards higher masses will require the design and development of smaller cavities and higher frequency, low noise sensors.

\begin{figure}[t]
  \begin{center}
    \includegraphics[width=0.40\textwidth]{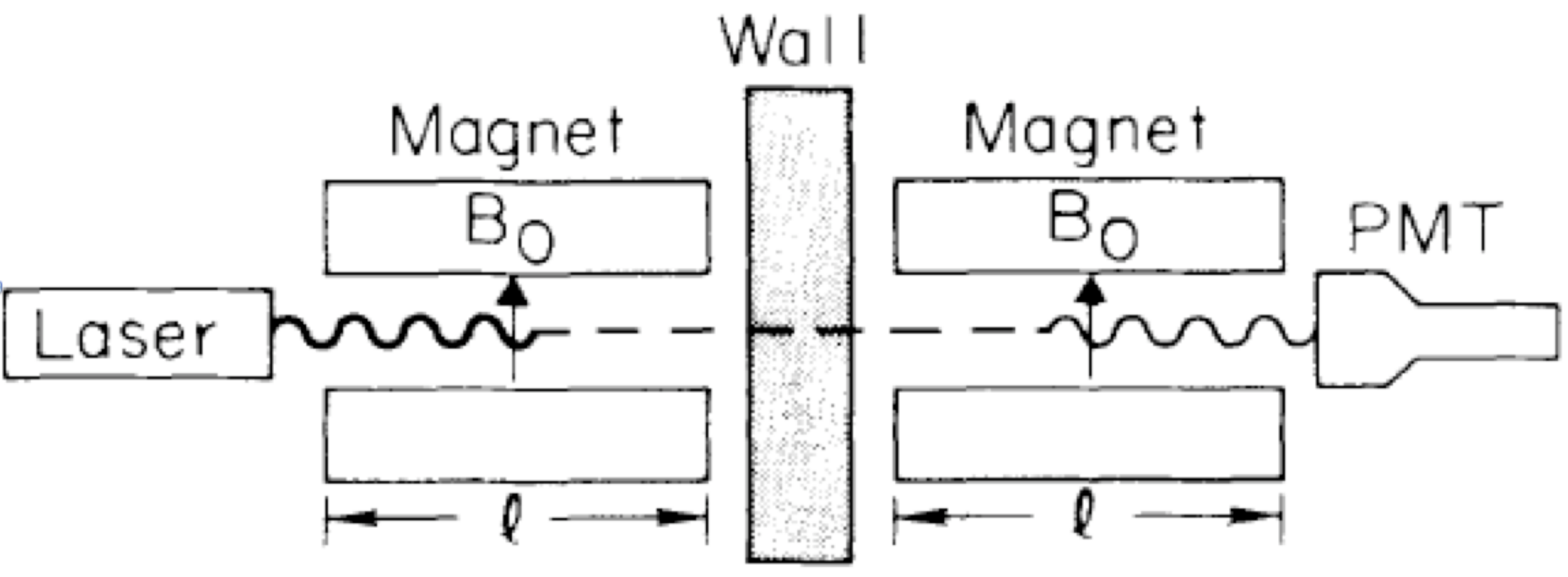}
  \end{center}
  \vspace{-1pc}
  \caption{In the presence of a background magnetic field, photons can oscillate into axions and vice-versa.  This leads to the phenomenon of light shining through walls, where photons can penetrate opaque regions in the guise of axions.  In the case of solar axion searches, the light source is the core of the sun, and the wall is the outer plasma layers of the sun.  In laser axion searches, the wall is a reflective mirror or beam dump.  For anomalous transparency observed from astrophysical sources, the wall is the bath of extragalactic background photons.}
\end{figure}

\section{Axion-photon oscillations}
The axion-photon interaction also manifests itself in the semiclassical equations of motion of a propagating axion-photon system.  Again, in the presence of a background magnetic field $B_0$, Eq.~\ref{E:interaction} becomes an off-diagonal term in a bilinear mass matrix, and induces transitions between photon flavor states (polarized along the direction of $B_0$) and axion flavor states.  In analogy with neutrino mixing, the transition probability is given by
\begin{equation}
\label{E:prob}
P_{\gamma\leftrightarrow a} = \sin^2(2\theta) \sin^2\left(\frac{(m_a^2-m_\gamma^2) L}{4 \omega}\right)
\end{equation}
where $m_a^2$ and $m_\gamma^2$ are the diagonal elements of the mass matrix, $\omega$ is the total energy of the wave, and the mixing angle $\theta$ is defined implicitly by the ratio of the off-diagonal elements to the diagonal elements
\begin{equation}
\label{E:theta}
\tan(2\theta) = \frac{2 g B_0 \omega}{m_a^2-m_\gamma^2}.
\end{equation}
In the limit of small mixing angle and magnetic lengths long compared to the natural oscillation length, 
\begin{equation}
\label{E:coh_limit}
L \ll L_\mathrm{osc} \equiv \frac{2 \omega}{m_a^2 - m_\gamma^2},
\end{equation}
the sines in Eq.~\ref{E:prob} can be replaced by their arguments, and the conversion probability simplifies to
\begin{equation}
\label{E:p_coh}
P_{\gamma\leftrightarrow a} \approx \frac{1}{4} (g B_0 L)^2.
\end{equation}
This is referred to as the coherent limit, in which the mixed amplitude monotonically grows with propagation distance.  Experiments have their maximum sensitivity to $g$ when operating in this limit.

\begin{figure}[t]
  \begin{center}
    \includegraphics[width=0.47\textwidth]{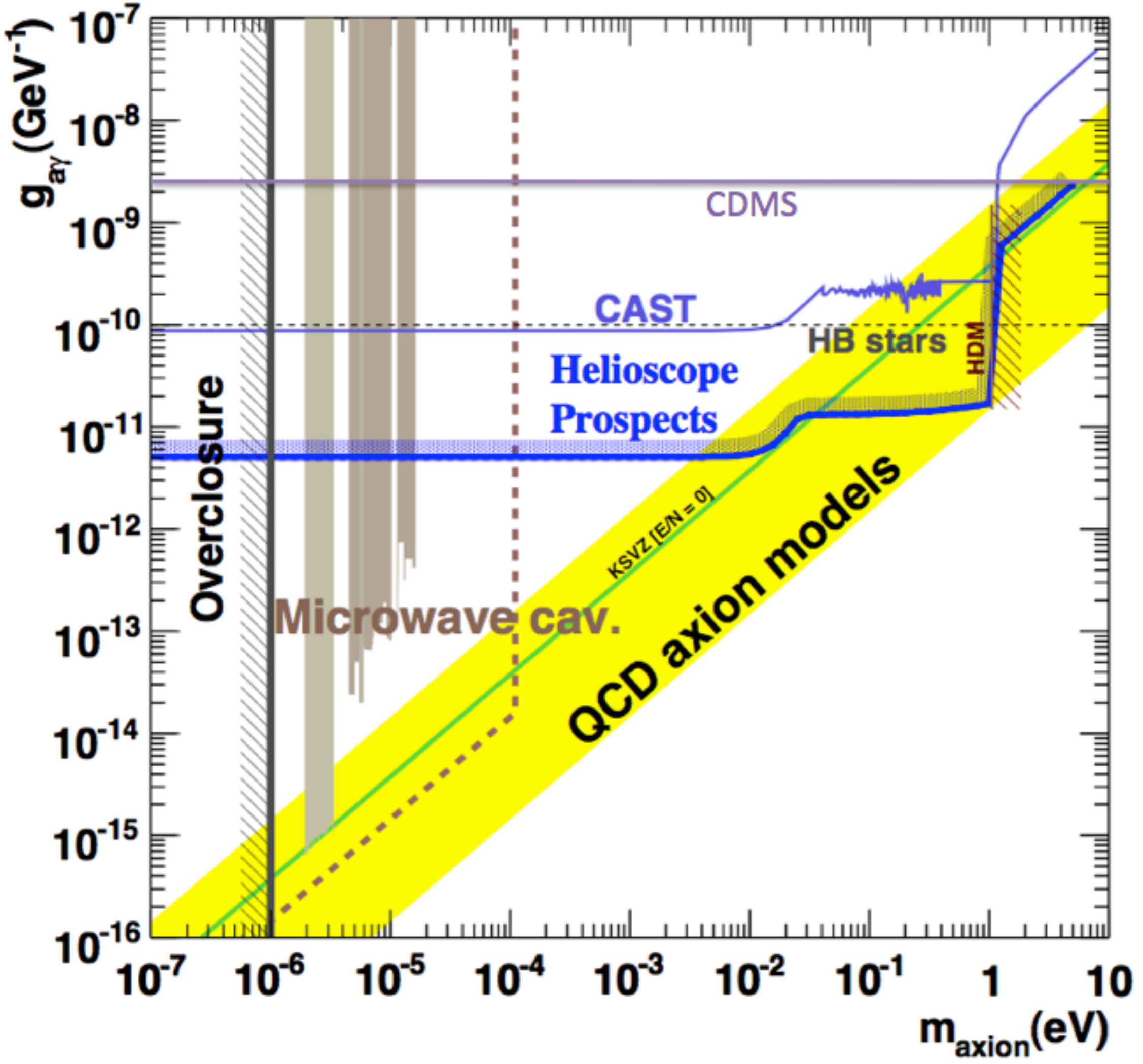}
  \end{center}
  \vspace{-1pc}
  \caption{Summary of experimental results from microwave cavity searches for dark matter axions (assuming axions saturate the dark matter density), helioscope searches from CAST, and Bragg scattering search from CDMS.  Helioscope prospects are bright as new magnets with $B_0 L \approx (14 T)(15 m)$ are developed for future accelerators projects.  Also shown is the upper bound on $g$ from star cooling of horizontal branch stars, the weak lower bound on $m_a$ from overproducing dark matter, and the diagonal band representing typical QCD axion models.}
\end{figure}

\section{Solar axion searches via oscillation}
Experiments such as the CERN Axion Solar Telescope (CAST) \cite{Arik:2008mq, Andriamonje:2007ew} or the Tokyo Axion Helioscope (Sumico) \cite{Inoue:2008zp} use the sun as a possible source of keV energy axions.  These axions can then be converted into detectable keV x-rays via oscillations within laboratory magnetic fields.  At larger values of $g$, axions would be copiously produced in the hot core of the sun by Primakoff scattering, using the axion-photon interaction vertex.  The axions would have a blackbody-like spectrum reflecting the temperature of the production region.  The mean free path for reconversion into photons is longer than the radius of the sun, and so these axions free stream out, contributing to the cooling of the sun.  The total axion flux is estimated to be 
$\Phi_a = 3.67\times 10^{11}\mbox{ cm}^{-2}\mbox{s}^{-1}(g/10^{-10}\mbox{ GeV}^{-1})^2$ and is proportional to $g^2$.  For $g > 10^{-10}\mbox{ GeV}^{-1}$, the cooling rate becomes inconsistent with the observed age of stars.  For the 9T, 9.3m accelerator dipole magnet with 14.5~cm$^2$ aperture used by the CAST experiment, the reconversion probability is
\begin{equation}
P_{a\rightarrow\gamma} = 1.7\times 10^{-17}\left(\frac{B_0 L}{9\mbox{T}\cdot 9.3\mbox{m}}\right)^2 \left(\frac{g}{10^{-10}\mbox{ GeV}^{-1}}\right)^2,
\end{equation}
indicating that sensitivity beyond the star cooling limit is possible with reasonable integration time.  The signal rate for coherent oscillation is given by
\begin{equation}
R_\mathrm{signal} = A \Phi_a P_{a\rightarrow\gamma} \propto A g^2 (g B_0 L)^2
\end{equation}
where $A$ is the collecting area.  A variety of x-ray detectors have been used, including PIN photodiodes by the Sumico experiment, and MicroMegas, TPCs, and x-ray telescopes to tightly focus the spot size on low noise CCDs.  All of these detectors have an intrinsic background rate $R_\mathrm{bkgd}$, and so in the absence of signal, the upper limit on the signal rate scales as the square root of integration time $t_\mathrm{int}$ and gives the design equation
\begin{equation}
g_\mathrm{limit} \propto \left(\frac{R_\mathrm{bkgd}}{t_\mathrm{int}}\right)^{1/8} \frac{1}{(B_0 L)^{1/2} A^{1/4}},
\end{equation}
applicable to searches for axions of sufficiently low mass to satisfy the coherence condition Eq.~\ref{E:coh_limit}.  The CAST experiment has achieved the world's best model-independent sensitivity to the photon coupling $g < 8.8\times 10^{-11}\mbox{ GeV}^{-1}$, which slightly extends the upper bound from star cooling arguments.

The current effort in both CAST and Sumico is to extend the coherent sensitivity to larger axion mass by increasing the effective mass of the photon to reduce the momentum difference between photons and axions and increase the coherent oscillation length $L_\mathrm{osc}$ in Eq.~\ref{E:coh_limit}.  A gas of neutral helium introduced into the magnet bore will look like a plasma to x-ray photons, which can resolve the individual electrons in the atoms.  The effective photon mass is then equal to the Debye frequency  
\begin{equation}
m_\gamma \equiv \sqrt{\frac{4\pi\alpha n_e}{m_e}} \approx  \sqrt{0.02 \frac{P[\mbox{mbar}]}{T[\mbox{K}]}} \ [\mbox{eV/c}^2]
\end{equation}
where $n_e$ is the electron number density, and $P$ and $T$ are the pressure and temperature of the gas.  Sumico uses a 4T, 2.3m magnet of 3.1 cm$^2$ aperture.  The $B\times L$ is an order of magnitude smaller than CAST, and so the detector is intrinsically less sensitive.  However, the shorter length allow Sumico to naturally probe towards larger axion masses while remaining the coherent limit.  Furthermore, the smaller magnetic field allows operation of the magnet at higher temperature $T\approx 6$~K which allows higher pressure levels of He$^4$ gas and correspondingly higher $m_\gamma$.  For magnetic field 9.3T, the CAST superconducting magnet must be operated at 1.8K at which increased pressure values would cause the He$^4$ gas to condense.  Instead, CAST must use the more expensive He$^{3}$ gas to reach higher pressure levels.  Both experiments aim to probe towards eV mass axions by performing a scan in steps of gas pressure to optimize the detector for finite bandwidth steps in axion mass.  Since each trial necessarily uses less integration time than the lower mass coherent data runs, the sensitivity to $g$ is somewhat poorer, and both experiments can only probe in a region of parameter space disfavored by star cooling, and by the SN1987A energy loss rate.  However, this region has never been directly tested using sources as well-modeled as the sun and surprises may be in store.

Future efforts on CAST will be focused in reducing detector background rates, and utilizing larger aperture, higher field accelerator magnets as they are developed for future accelerator projects.  Already, the background rates of the MicroMegas detectors have been improved by an order of magnitude by using better shielding, and low radioactivity materials.  If 14 T, 15 m magnets become available it may be possible for an upgraded CAST experiment to achieve sensitivity at the level of $g \approx 5\times 10^{-12}\mbox{ GeV}^{-1}$.

\section{Solar search via Bragg scattering}
Similarly, a search for solar axions has been conducted using the Cryogenic Dark Matter Search (CDMS) detectors \cite{Collaboration:2009ht}.  The detection technique is now Bragg-like scattering in which the amplitudes for Primakoff scattering from individual electrons in an ordered Germanium crystal lattice add coherently for orientations of the beam and the crystal lattice which satisfy the usual Bragg condition.  The coherence enhances the scattering probability by factors up to $\sim 1000$.    In this scattering process, the axion is converted into an x-ray photon which immediately deposits energy into the crystal in the form of ionization and heat.  Nuclear recoil events can be rejected using the ratio of ionization and phonon signals, using the standard CDMS WIMP search technique.  However, these axion-induced $\gamma$ events cannot be distinguished from other $\beta$ or $\gamma$ events on the basis of signal risetime.  A background rate of $\sim 1.5$~count/day/kg/keV from electromagnetic events in both the surface and bulk of the crystals is seen, in an analysis window of 2.5-8 keV energy, as measured by the phonon signal.  To focus the search for an axion beam from the sun, a maximum likelihood analysis is performed to look for excess event rates as a function of measured energy and time of day, which determines the instantaneous deviation from the Bragg angle as a function of energy.  No excess was detected in a 443.2 kg-day sample, resulting in an upper limit $g<2.4\times 10^{-9}\mbox{ GeV}^{-1}$, independently of axion mass up until the kinematic limit $m_a< T_\mathrm{sun} \approx 10^4$~K.  Development of lower background detectors such as a solid xenon crystal, are underway and could further improve this limit by an order of magnitude.

\section{Laser axion searches} 
Several experiments have also been conducted using lasers as axion sources.  The laser beam is directed into a transverse magnetic field $B_0$ in which an axion beam is generated by the photon-axion oscillation process.  The preferential depletion of one component of polarization with respect to the $B_0$ direction can manifest itself as a rotation of the net polarization of the laser beam.  An anomalous detection of such polarization rotation was published but later retracted by the PVLAS experiment \cite{Zavattini:2005tm}.  The inferred coupling was unexpectedly large $g\approx 2\times 10^{-6}\mbox{ GeV}^{-1}$, and inspired the development of a number of new experiments (GammeV, BMV, LIPSS, ALPS, OSQAR), to verify this signal.  

All of these experiments are based on the idea of shining light through walls.  A laser beam is sent into a transverse magnetic field where a tiny fraction of the laser photons is converted into axions.  An opaque wall is inserted into the laser path to block or deflect the unconverted photons.  Being very weakly-interacting, the axion beam passes unhindered through the wall and into a second magnetic field region beyond where they oscillate back into photons.  The regenerated photons, which have apparently passed right through the wall, can be detected with a low noise photodetector.  The probability for photon oscillation and reconversion has two factors of Eq.~\ref{E:p_coh}
\begin{equation}
P_\mathrm{regen} = \frac{1}{16} (g B_0 L)^4,
\end{equation}
assuming equal $B_0 L$ on both sides of the wall.  For background-free single photon detection, the sensitivity to low rates scales linearly with integration time $t_\mathrm{int}$, and the sensitivity to the coupling scales as
\begin{equation}
\label{E:regen_scaling}
g_\mathrm{limit} \propto \frac{1}{B_0 L t_\mathrm{int}^{1/4}}.
\end{equation}   
If background subtraction is required, then the sensitivity to $g$ scales as $t_\mathrm{int}^{-1/8}$.

As an example, the GammeV experiment \cite{Chou:2007zzc} has achieved nearly background-free operation by using a pulsed 1064 nm laser which produced 20 Hz of 5 ns pulses which went into a 5T dipole magnet from the Tevatron accelerator.  A beam-blocking mirror is inserted into the magnet bore to divide the 6m length into axion production and photon regeneration regions, each of which is contained in a separately isolated vacuum system.  Using timing circuits with $\sim 1$~ns precision, the search for regenerated photons can be performed in 10 ns windows in coincidence with the pulse arrival times.  The background $\sim 100$~Hz dark rate of a cooled photomultiplier tube the produces an accidental coincidence rate of only $2\times 10^{-5}$~Hz and allows 15 hours of background free running.  No excess signal has been seen, resulting in a limit of $g < 3.5\times 10^{-7}\mbox{ GeV}^{-1}$.  This limit for odd-parity particles, and a similar limit achieved for even-parity particles achieved by rotating the initial laser polarization by $90^\circ$, completely rule out the explanation for the PVLAS anomaly in terms of axion-like particles.  These results are the current best laser limits on the axion-photon coupling, but are unfortunately still far from the astrophysical limits from star cooling and solar searches.  Further sensitivity to $g$ via this technique is limited by the availability of stronger and longer magnets.

\begin{figure}[t]
  \begin{center}
    \includegraphics[width=0.47\textwidth]{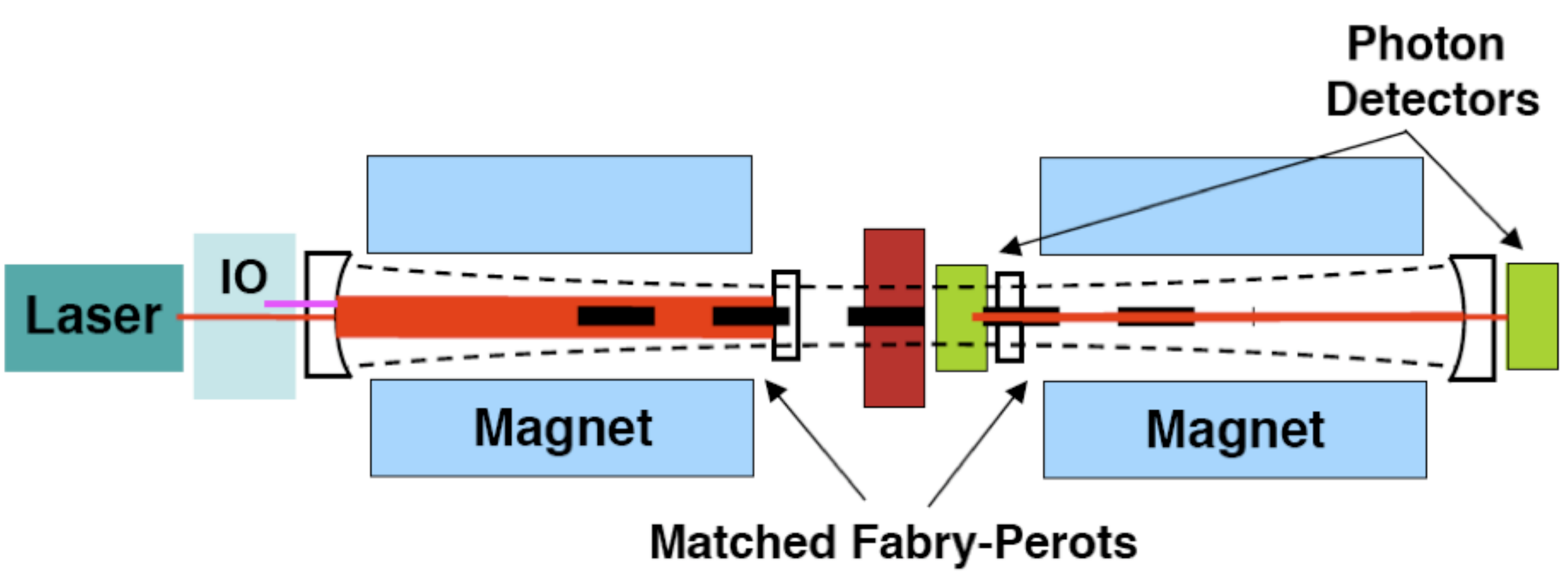}
  \end{center}
  \vspace{-1pc}
  \caption{By placing an optical cavity around the axion generation region, the initial laser beam can be recycled and passed many times through the magnetic field.  The produced axion beam is also shaped by the curvature of the mirrors.  A second optical cavity is placed around the photon regeneration region, with curvature matched to the divergence of the axion beam.  Regenerated standing waves can then build up coherently over the lifetime of the cavity, resulting in a large enhancement in the number of regenerated photons.}
\end{figure}

\section{Cavity-enhanced photon regeneration}
An ingenious proposal has recently been made to coherently enhance the photon regeneration rate using high finesse Fabry-Perot optical cavities \cite{Sikivie:2007qm, Mueller:2009wt}.  A cavity is placed around the axion-generating magnet in order to recycle the photon beam, or equivalently to increase the instantaneous power within the cavity by a factor of $\mathcal{F}/\pi$.  The finesse $\mathcal{F}$ is roughly equal to the number of mirror reflections a photon makes before it escapes from the cavity.  The curvature of the cavity mirrors also determines the transverse Gaussian profile of the trapped standing wave.  As long as the momentum difference between axions and photons is small, $\Delta m^2/2\omega \ll \omega$, the produced axion beam will have approximately the same transverse momentum distribution as the photon beam.  The resulting Gaussian axion beam will undergo diffraction-limited expansion as it leaves the cavity and passes through an opaque wall.  Beyond this wall, a second cavity is placed around the photon-regenerating magnet.  The curvature of the mirrors is matched to the shape of the expanding axion beam so that any regenerated electromagnetic field can efficiently populate a standing wave in the second cavity.  If the relative phase between the two cavities can be maintained, then the standing wave electric field will build up coherently in the second cavity, again over approximately $\mathcal{F}$ reflections and gives an enhancement of $\mathcal{F}^2$ in photon number.  In effect, axions generated from multiple reflections in the first cavity all contribute coherently to the photon standing wave in the second cavity, resulting in this large ``wavefunction-squared'' enhancement in the total cavity power.   The regenerated photon signal can then be seen as the regenerated photons leak out through the end mirror.   If the cavity losses which determine $\mathcal{F}$ are dominated by transmission through this end mirror, then the instantaneous leaked photon rate is suppressed by a factor of $1/\mathcal{F}$.  The net result is that the regenerated photon rate is coherently enhanced by a factor of $\mathcal{F}^2$, potentially an extremely large number.  For a laser photon rate $R_\mathrm{laser}$, the number of detectable regerated photons is  
\begin{equation}
N_\mathrm{regen} = (R_\mathrm{laser} t_\mathrm{int}) \frac{1}{16} (g B_0 L)^4 \left(\frac{\mathcal{F}}{\pi}\right)^2.
\end{equation}

Optical heterodyne detection can be used to sense the regenerated field which has been transmitted out from the second cavity.  A secondary laser is phase-locked to the axion-generating laser, but at a fixed, stable frequency offset, $\omega_2 = \omega_1 + \Delta\omega$.  The regenerated photons at frequency $\omega_1$ are focused onto a photodiode where they overlap with the beam from the second laser which serves as the local oscillator (LO).  The RF beat frequency then has power proportional to $\sqrt{R_\mathrm{regen} R_\mathrm{LO}}$.  In this way, the photon flux $R_\mathrm{LO}$ of the LO laser amplifies the regenerated photon signal.  The amplification factor can be made almost arbitrarily large until the shot noise of the LO laser dominates the total noise of the detection system.  In this shot-noise-limited regime, a heterodyne detection system has single-photon sensitivity, and quantum-limited noise performance.  The noise originates from the zero-point fluctuations of the states being mixed to produce the beat signal.  It obeys an exact upper bound of 2 photons per integration time, and does not have a Poisson tail.  Subtraction of a constant Poisson background rate is therefore not necessary, and the sensitivity of the technique scales as   
\begin{equation}
g_\mathrm{limit}\propto \frac{1}{(R_\mathrm{laser} t_\mathrm{int})^{1/4} (B_0 L) (\mathcal{F})^{1/2}}.
\end{equation}

Several experimental groups around the world are exploring an implementation of the cavity-enhanced photon regeneration technique.  As an example, a proposed Fermilab experiment would use a configuration of 6+6 Tevatron dipole magnets, each operating at 5 T, giving a 36 m baseline for each conversion region.  With $\mathcal{F}\approx 10^5$, this experiment would have sensitivity to $g\approx 2\times 10^{-11}\mbox{ GeV}^{-1}$.  Just as with solar searches, this sensitivity can be improved with higher field magnets which may become available in the future.  The magnetic length can also be extended up to several hundred meters before the $\sim 10$~cm aperture diameter of typical accelerator magnets starts to clip the Gaussian cavity modes.  With these future devices, it may be possible to reach sensitivity to photon couplings as low as $g\approx 6\times 10^{-13}\mbox{ GeV}^{-1}$.

\begin{figure}[t]
  \begin{center}
    \includegraphics[width=0.49\textwidth]{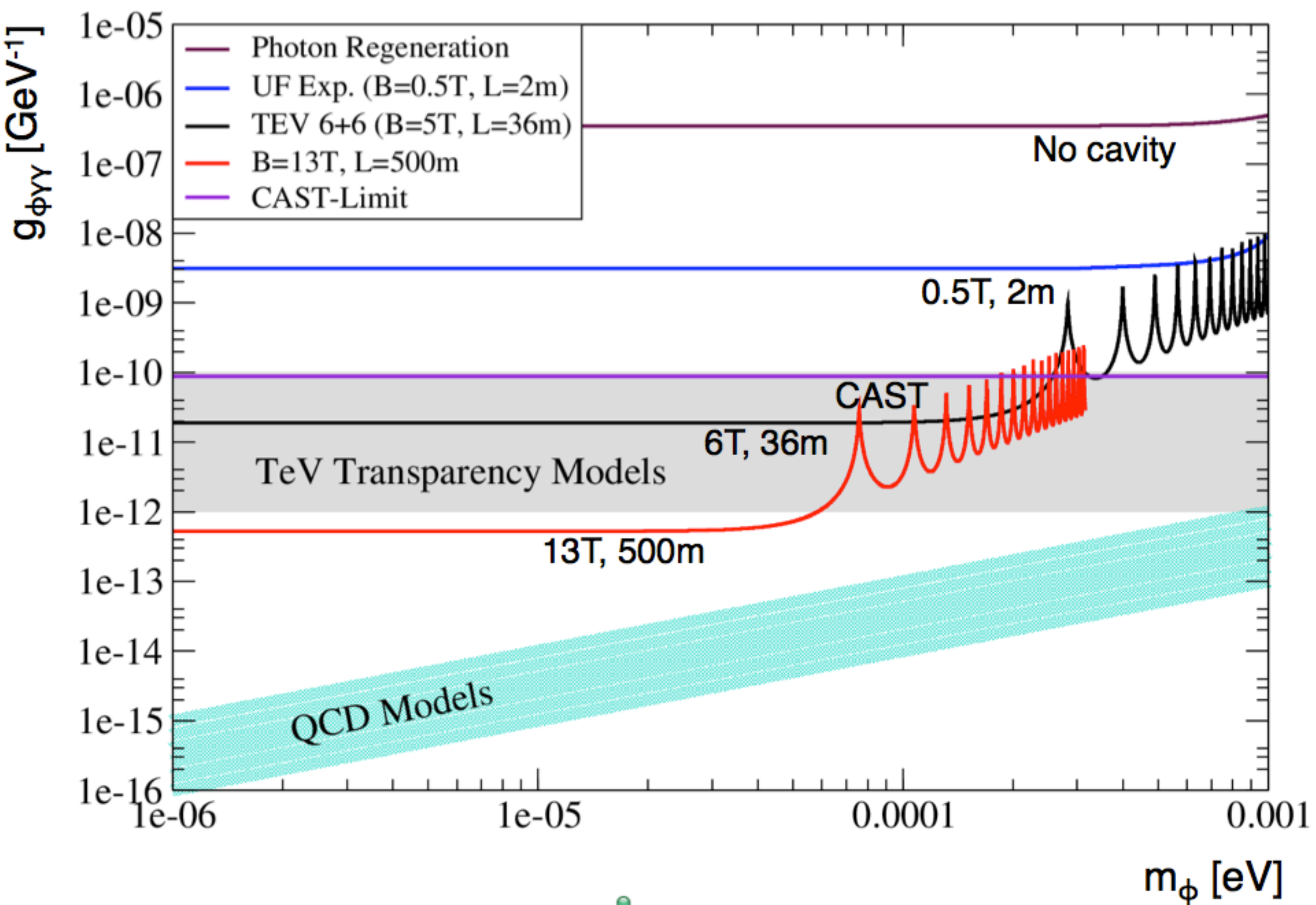}
  \end{center}
  \vspace{-1pc}
  \caption{The new idea of using optical cavities to coherently enhance the photon regeneration rate will improve laser search sensitivity by several orders of magnitude to cover a region of parameter space suggested by recent astrophysical observations of anomalous transparency of the universe to high energy photons.  Shown here are the laser limits of a no-cavity configuration from GammeV, compared to the gains in sensitivity that could be made with cavities of finesse $\mathcal{F}=10^4-10^5$, combined with longer magnetic baselines $B_0 L$.}
\end{figure}

\section{Extragalactic axions}
Recent astrophysical observations indicate the possibility of a new axion-like particle with photon coupling $g\approx 10^{-11}\mbox{ GeV}^{-1}$ within the reach of  the next generation of oscillation experiments.  The universe is opaque to high energy photons which can pair-produce on the gas of photons composing the extragalactic background light (EBL).  The EBL consists of photons from all sources including the cosmic microwave background and redshifted starlight from the first stars.  Anomalous transparency has been reported in observations by various air Cherenkov telescopes of TeV photons from distant blazars \cite{Aharonian:2005gh}.  TeV photons can scatter efficiently on the near-infrared background photons of micron wavelength with interaction length of several hundred Mpc, given preliminary estimates of the EBL spectral density.  However, the spectra of TeV blazars at distanes of hundreds of Mpc do not exhibit the expected exponential suppression in flux.  The lack of flux suppression in most blazar observations has been used as a tool to directly constrain the EBL spectral density to below the levels implied by direct satellite observations.  These satellite observations may indeed be problematic due to possible contamination from isotropically scattered sunlight with the solar system.

However, another explanation for the perhaps anomalous transparency of the universe to TeV gamma rays is that TeV photons may be converted to energetic axions in the magnetic fields of the astrophysical sources.  The axions may then shine through the ``wall'' of extragalactic background light before oscillating back into photons in the galactic magnetic field.  Because photon-axion oscillations only affect a single polarization component, the lack of obvious spectral features in the TeV spectra, due to the attenuation of the unconverted component, then requires very efficient conversion and regeneration of photons.  Recalling Eq.~\ref{E:p_coh}, $\mathcal{O}(1)$ probabilities requires $g\approx 1/(B_0 L)$.  Observations from the Pierre Auger Observatory indicate that ultra-high-energy cosmic rays are very likely protons which are accelerated in extragalactic astrophysical sources \cite{Cronin:2007zz}.  In order to contain the protons while accelerating them, the sources must have magnetic field regions which satisfy $B_0 L = E_\mathrm{CR} \approx 10^{20}\mbox{ eV}$ for the highest energy cosmic rays.  In a somewhat miraculous coincidence, some models of the galactic magnetic field also give a comparable magnetic baseline for the poloidal magnetic field \cite{Harari:1999it}, $B_0 L \approx (6 \mu\mbox{G})\cdot (4\mbox{ kpc}) = 3\times 10^{19}\mbox{ eV}$.  So an axion model with $g\approx 10^{-11}\mbox{ GeV}^{-1}$ would predict efficient conversions via oscillation both at the source, and in the galaxy \cite{Hooper:2007bq,Simet:2007sa}.  This mechanism is expected to be effective at photon energies up to $1$~TeV, above which magnetic birefringence due to quantum electrodynamics sufficiently alters the photon dispersion relationship such that the effective photon-axion mixing angle becomes small \cite{Hochmuth:2007hk}.  An alternative model has photons oscillating into axions in intergalactic magnetic domains of nG field strength and Mpc scale \cite{DeAngelis:2007dy}.

\section{Cosmic Microwave Background observations} 
CMB observations have also been used to constrain the joint axion and inflaton parameter space \cite{Hamann:2009yf}.  The results depend on whether the axion symmetry breaking scale $f$ is greater than or less then the Hubble rate during inflation $H_I$.  If $f<H_I$ then the Peccei-Quinn symmetry breaks after inflation, leading to the formation of topological axionic strings.  This scenario is more likely for larger values $H_I>10^{14}$~GeV in which the primordial spectrum of gravity waves may be observable via their imprint of B-mode polarization patterns on the CMB.  In this case, determination of $H_I$ via CMB polarization places an upper bound on $f$.
 
However, if $f>H_I$, then the entire cosmological horizon contained a single mean misalignment angle $\Theta$ from which the Goldstone boson field started rolling during the QCD phase transition.  This misalignment angle determines what fraction of energy from the quark-gluon plasma is converted into axionic potential energy, and ultimately into cold dark matter axions.  However, scalar perturbations of the Goldstone boson field due to inflation generate local deviations in $\Theta$ with Gaussian variance $\delta \Theta^2 \propto (H_I/f)^2$.  Because these perturbations do not inject more energy, but instead determine how the local energy density is partitioned, they give rise to isocurvature perturbations on the CMB spectrum.  The CMB spectrum, measured most recently by WMAP, is most consistent with a spectrum of adiabatic perturbations with very little isocurvature component.  The CMB spectrum therefore places an upper bound on $\delta \Theta^2$ and hence on the parameter space of $f$ and $H_I$.  A positive detection of isocurvature would imply a fairly low $H_I<10^{8}$~GeV for natural models of axion dark matter with $f < 10^{13}$~GeV.  If isocurvature measurements are ambiguous, then determination of $f$ via direct axion dark matter searches would exclude a range of $H_I$.

\begin{figure}[t]
  \begin{center}
    \includegraphics[width=0.49\textwidth]{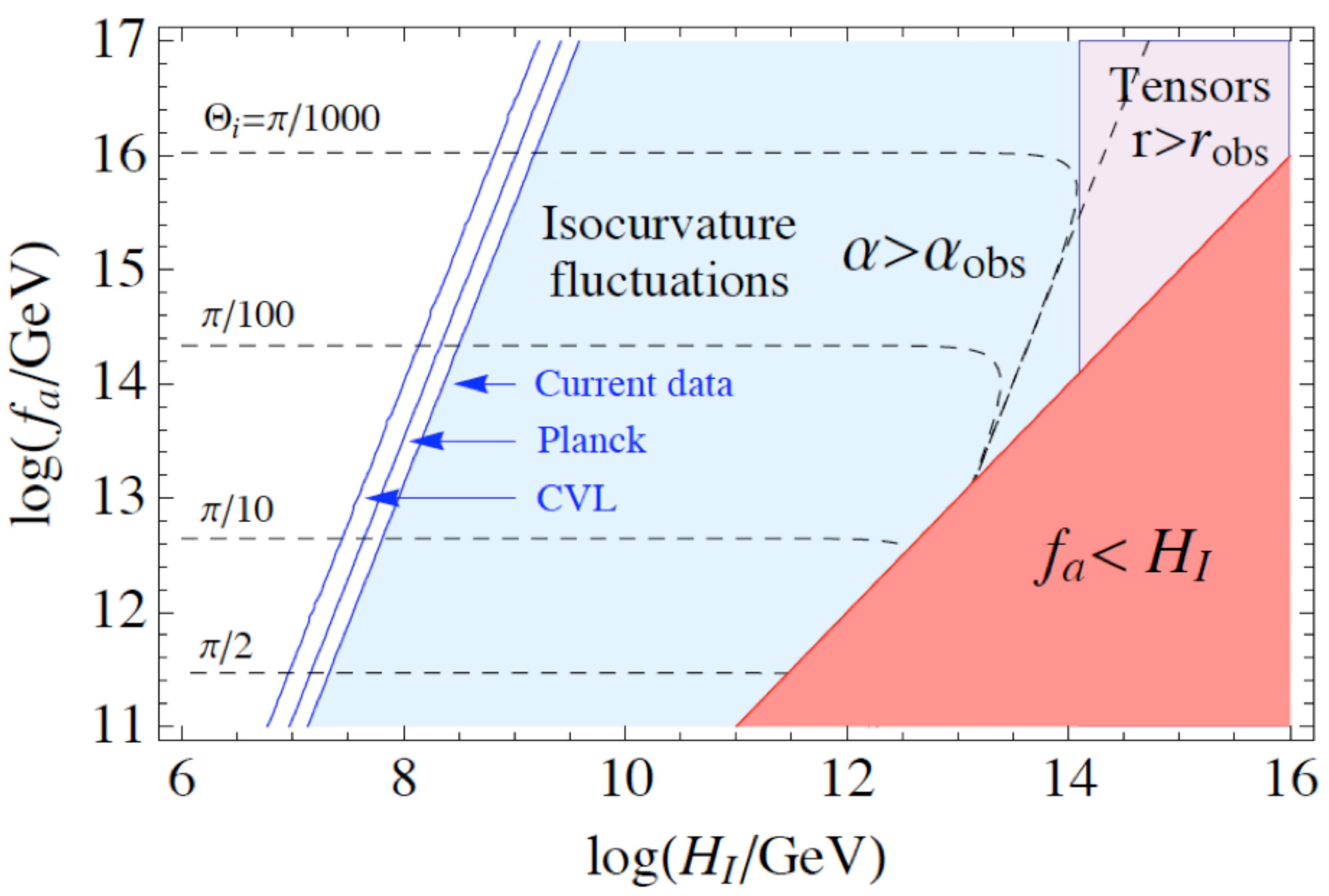}
  \end{center}
  \vspace{-1pc}
  \caption{Constraints from WMAP data on the joint parameter space of Peccei-Quinn scale $f$ and the inflation scale $H_I$.  Upper bounds on the tensor-to-scalar ratio and of isocurvature component of the CMB power spectrum constrain this parameter space. }
\end{figure}

\section{Summary} 

The outlook for the experimental axion field is bright.  A variety of techniques have been proposed and implemented, using semiclassical methods to overcome the suppression of quantum amplitudes by the large Peccei-Quinn scale.  Recent breakthroughs in sensor technology, magnet development, and experimental configuration, promise order of magnitude improvements in sensitivity to the photon-axion coupling constant.  In particular, a new idea of integration of optical cavities into laser axion searches will produce several orders of magnitude improvement in sensitivity.   Hints of axion-like particles are seen in astroparticle data, and will be checked with laboratory probes.  Ideas about the role of axions in cosmology are being refined and checked against new data.

As a side benefit, data from axion search experiments is also being used to constrain many types of models of other low-mass particles such as chameleons or paraphotons.  The discussion of this fascinating work is beyond the scope of this paper but can be found elsewhere.    

This work is supported by the U.S. Department of Energy under contract No. DE-AC02-07CH11359.

                       \vspace{-.5\baselineskip}




\begin{thebibliography}{99}
\bibitem{Peccei:1977ur}
  R.~D.~Peccei and H.~R.~Quinn,
  Phys.\ Rev.\  D {\bf 16}, 1791 (1977).

\bibitem{Peccei:1977hh}
  R.~D.~Peccei and H.~R.~Quinn,
  Phys.\ Rev.\ Lett.\  {\bf 38}, 1440 (1977).

\bibitem{Wilczek:1977pj}
  F.~Wilczek,
  Phys.\ Rev.\ Lett.\  {\bf 40}, 279 (1978).

\bibitem{Weinberg:1977ma}
  S.~Weinberg,
  Phys.\ Rev.\ Lett.\  {\bf 40}, 223 (1978).

\bibitem{Duffy:2006aa}
  L.~D.~Duffy {\it et al.},
  Phys.\ Rev.\  D {\bf 74}, 012006 (2006)
  [arXiv:astro-ph/0603108].

\bibitem{Duffy:2005ab}
  L.~Duffy {\it et al.},
  Phys.\ Rev.\ Lett.\  {\bf 95}, 091304 (2005)
  [arXiv:astro-ph/0505237].

\bibitem{Bradley:2003kg}
  R.~Bradley {\it et al.},
  Rev.\ Mod.\ Phys.\  {\bf 75}, 777 (2003).

\bibitem{Arik:2008mq}
  E.~Arik {\it et al.}  [CAST Collaboration],
  JCAP {\bf 0902}, 008 (2009)
  [arXiv:0810.4482 [hep-ex]].

\bibitem{Andriamonje:2007ew}
  S.~Andriamonje {\it et al.}  [CAST Collaboration],
  JCAP {\bf 0704}, 010 (2007)
  [arXiv:hep-ex/0702006].

\bibitem{Inoue:2008zp}
  Y.~Inoue, Y.~Akimoto, R.~Ohta, T.~Mizumoto, A.~Yamamoto and M.~Minowa,
  Phys.\ Lett.\  B {\bf 668}, 93 (2008)
  [arXiv:0806.2230 [astro-ph]].

\bibitem{Collaboration:2009ht}
  Z.~Ahmed {\it et al.}  [CDMS Collaboration],
  arXiv:0902.4693 [hep-ex].

\bibitem{Zavattini:2005tm}
  E.~Zavattini {\it et al.}  [PVLAS Collaboration],
  Phys.\ Rev.\ Lett.\  {\bf 96}, 110406 (2006)
  [Erratum-ibid.\  {\bf 99}, 129901 (2007)]
  [arXiv:hep-ex/0507107].

\bibitem{Chou:2007zzc}
  A.~S.~..~Chou {\it et al.}  [GammeV (T-969) Collaboration],
  Phys.\ Rev.\ Lett.\  {\bf 100}, 080402 (2008)
  [arXiv:0710.3783 [hep-ex]].

\bibitem{Sikivie:2007qm}
  P.~Sikivie, D.~B.~Tanner and K.~van Bibber,
  Phys.\ Rev.\ Lett.\  {\bf 98}, 172002 (2007)
  [arXiv:hep-ph/0701198].

\bibitem{Mueller:2009wt}
  G.~Mueller, P.~Sikivie, D.~B.~Tanner and K.~van Bibber,
  arXiv:0907.5387 [hep-ph].

\bibitem{Aharonian:2005gh}
  F.~Aharonian {\it et al.}  [H.E.S.S. Collaboration],
  Nature {\bf 440}, 1018 (2006)
  [arXiv:astro-ph/0508073].

\bibitem{Cronin:2007zz}
  J.~Abraham {\it et al.}  [Pierre Auger Collaboration],
  Science {\bf 318}, 938 (2007)
  [arXiv:0711.2256 [astro-ph]].

\bibitem{Harari:1999it}
  D.~Harari, S.~Mollerach and E.~Roulet,
  JHEP {\bf 9908}, 022 (1999)
  [arXiv:astro-ph/9906309].

\bibitem{Hooper:2007bq}
  D.~Hooper and P.~D.~Serpico,
  Phys.\ Rev.\ Lett.\  {\bf 99}, 231102 (2007)
  [arXiv:0706.3203 [hep-ph]].

\bibitem{Simet:2007sa}
  M.~Simet, D.~Hooper and P.~D.~Serpico,
  Phys.\ Rev.\  D {\bf 77}, 063001 (2008)
  [arXiv:0712.2825 [astro-ph]].



\bibitem{Hochmuth:2007hk}
  K.~A.~Hochmuth and G.~Sigl,
  Phys.\ Rev.\  D {\bf 76}, 123011 (2007)
  [arXiv:0708.1144 [astro-ph]].

\bibitem{DeAngelis:2007dy}
  A.~De Angelis, O.~Mansutti and M.~Roncadelli,
  Phys.\ Rev.\  D {\bf 76}, 121301 (2007)
  [arXiv:0707.4312 [astro-ph]].

\bibitem{Hamann:2009yf}
  J.~Hamann, S.~Hannestad, G.~G.~Raffelt and Y.~Y.~Y.~Wong,
  JCAP {\bf 0906}, 022 (2009)
  [arXiv:0904.0647 [hep-ph]].

\end{thebibliography}
\end{document}